\documentclass{article}
\newcommand{\bfr}{\begin{flushright}}
\newcommand{\efr}{\end{flushright}}
 
\begin{document}
\title{The Friedmann Universe and Compact Internal Spaces in
Higher-Dimensional Gravity Theories
%\thanks{Presented at ...}%
}
\author{Kiyoshi Shiraishi\\
%\address{
Department of Physics, Tokyo Metropolitan University,
Tokyo 158%}
}
\date{Prog. Theor. Phys. Vol. 76, No. 1, July 1986, pp. 321--324,
Progress Letters
}
\maketitle
\begin{abstract}
We consider gravity theories in $4+N$ dimensions which are governed by
the Lagrangian written as an extended Gauss-Bonnet density. We can find
a naturally generalized Einstein gravity where the maximal symmetric
compactification leads to vanishing four-dimensional cosmological
constant in the static limit. A later stage in the generalized
Kaluza-Klein cosmology is also examined.
\end{abstract}
%\PACS{}

In the last decade, the unification of the gauge interactions and the
gravity through higher dimensions has much interest.\cite{1} This
renewed Kaluza-Klein idea is originated from the study of
supergravity theories\cite{2} and superstring theories,\cite{3}
which have a simple or unique structure in higher dimensions than four.

Most attractive feature of Kaluza-Klein theories is that the isometry
group of compact extra spaces will be viewed as the gauge symmetry group
in the four-dimensional effective theory.\cite{1} Various gauge groups
can be obtained as a consequence of the corresponding compactification
in the Kaluza-Klein (super) gravity.\cite{2}

Another aspect of the Kaluza-Kein model is the cosmological evolution
of the scale factors. In the early universe, it is natural to expect
the ``dimensional reduction transition''\cite{4} which explains the
reason why the length scale of extra spaces is too small to be
obervable. And also, we can ask whether the cosmological inflation
associated with this phase transition takes place or not.\cite{5}
However, before such investigations, we have to explain why the
four-dimensional cosmological constant is so small in
our universe.\cite{6} From the cosmological observation it follows that
the cosmological constant in the universe at present cannot be greater
than the critical density $\sim 10^{-120} m_{\rm pl}^4$. 
In usual Kaluza-Klein supergravity theories, though the
higher-dimensional cosmological term is forbidden by supersymmetry, the
compactification of extra spaces brings about a large four-dimensional
cosmological constant $\sim m_{\rm pl}^4$. On the other hand, in
non-super-theories, we can adjust the higher-dimensional cosmological
constant in order to lead a vanishing four-dimensional cosmological
constant. However, this fine-tuning seems to be unnatural while we
have no principle to decide any fixed values for higher-dimensional
cosmological constant.

There are many attempts to solve this ``cosmological constant
problem''.  Wetterich and his coworkers adopted non-compact spaces
as extra spaces, and showed that no fine-tuning is needed to obtain a
vanishing four-dimensional cosmological constant.\cite{7} On the other
hand, in the low energy limit of the superstring theories, it is
favoured that extra six dimensions are compactified on the Calabi-Yau
manifolds, Ricci curvature of which is zero.\cite{8} In both cases, the
compactification yields few or zero numbers of Kaluza-Klein gauge
fields, because the symmetries of the internal spaces are far from
maximal. Thus the ``beautiful concept'' of
Kaluza-Klein theories is badly spoiled.

Another approach to the cosmological constant problem is given by
Gasperini.\cite{9} He demonstrated that an induced gravity model of
Zee's type\cite{10} with another matter field in higher dimensions
solves the problem. But in this case, the choice
of the matter fields is very crucial.

In this paper, we respect simplicity. We abandon the Einstein-Hilbert
action in higher dimensions and, instead, we consider a generalized
pure gravity action. First of all, we look for symmetric tensors with
the following properties:\cite{11}

(1) It is a concomitant of the metric tensor and its first two
derivatives.

(2) It is divergence free.

\noindent
When we set such a tensor equals zero, it can be regarded as the
gravitational fIeld equation in vacua.

In four dimensions, the tensor which satisfies the above condition is
only a linear combination of the Einstein tensor and the metric tensor.
But in higher dimensions, other tensors with these properties have been
found.\cite{11}  Recently, it is well mentioned that the Lagrangian
which leads to such a field equation is expressed in a linear
combination of extended Gauss-Bonnet densities.\cite{12,13} We have
partly motivated the generalized gravity theory based on such a
Lagrangian with the recent results on the low energy limit of string
theories.\cite{14} However, according to more recent investigations, it
seems that there is no exact Gauss-Bonnet form of pure gravity in the
effective field theory of strings.\cite{15,16}

In the present paper, we consider gravity theories in
$(4+N)$ dimensions, Lagrangian of which is a monomial of the
Gauss-Bonnet density, and investigate  the  maximal 
symmetric compactification of $N$ dimensions. We also investigate the
later stage of evolutions of scale factors in the model where a
vanishing four dimensional cosmological constant appears naively after
the compactification.

We consider $D=4+N$-dimensional space-time. Let $e^A$ be an orthonormal
basis for the metric $ds^2$:
\begin{equation}
ds^2=e^A\otimes e^B \eta_{AB}\,,\quad 
\eta={\rm diag}(-1,1,\cdots,1)\,.
\end{equation}
It is convenient to introduce the differential forms,
\begin{equation}
\varepsilon_{A_1\cdots A_m}=\frac{1}{(D-m)!}
\varepsilon_{A_1\cdots A_mA_{m+1}\cdots A_D} e^{A_{m+1}}
\wedge\cdots\wedge e^{A_D}
\end{equation}
where the Levi-Civita symbol is totally antisymmetric
with $\varepsilon_{1\cdots D}=1$. The connection one-form $\omega_{AB}$
is defined by
\begin{equation}
de^A+\omega^A{}_B\wedge e^B=0\,,\quad \omega_{AB}=-\omega_{BA}\,.
\end{equation}
(Here we only consider torsion-free theories.)
The curvature two-form $\Theta_{AB}$ related to the
Riemann tensor $R_{ABCD}$ by
\begin{equation}
\Theta_{AB}=\frac{1}{2} R_{ABCD} e^C\wedge e^D\,,
\end{equation}
is given by $\Theta^A{}_{B}=d\omega^A{}_B+\omega^A{}_C\wedge
\omega^C{}_B$.

The Lagrangian to be considered here is
\begin{equation}
{\cal L}=K {\cal L}_m\,,
\end{equation}
where $K$ is a constant and
\begin{equation}
{\cal L}_m=\Theta^{A_1B_1}\wedge\cdots\wedge\Theta^{A_mB_m}\wedge
\varepsilon_{A_1B_1\cdots A_mB_m}\,,\quad (2m\le D)
\end{equation}

This is the so-called ``dimensionally-extended Gauss-Bonnet
density.''\cite{12,13,14} In particular, we can see
\begin{eqnarray}
{\cal L}_0&=& e\,,\\
{\cal L}_1&=& \Theta^{AB}\wedge\varepsilon_{AB}=e
R\,,\\ {\cal L}_2&=&
\Theta^{AB}\wedge\Theta^{CD}\wedge\varepsilon_{ABCD}\,,\nonumber
\\ &=& e
(R_{ABCD}^2-4 R_{AB}^2+R^2)\,,\quad {\rm etc.}\,,
\end{eqnarray}
where $R_{AB}$ is the Ricci tensor, $R_{AB}=R^C{}_{ACB}$ and
$R$ is the scalar curvature, $R=R_A^A$. For later convenience, we define
$L_m$ as $eL_m ={\cal L}_m$. 

Now we consider the
compactification. First of all, we consider the case
\begin{equation}
e^\alpha=e^\alpha(x)\,,\quad e^a=e^a(y)\,,
\end{equation}
where $e^\alpha$ is a vierbein in four-dimensional spacetime, $e^a$ is a
vierbein in the $N$-dimensional internal space, and $x^\mu$ and $y^m$
represent four-dimensional and $N$-dimensional coordinates respectively.
As we will see later, this case includes the static compactification.

The curvature splits according to
\begin{equation}
\Theta^{AB}=\left\{
\begin{array}{c}
\Theta^{\alpha\beta}(x)\,,\\
\Theta^{ab}(y)\,,
\end{array}
\right.
\end{equation}
then it can be shown that \cite{12}
\begin{equation}
L_m=\sum_{\nu=0}^m\left(
\begin{array}{c}
m\\\nu
\end{array}
\right) \hat{L}_{m-\nu}\tilde{L}_\nu\,,
\end{equation}
where a roof refers to four-dimensional spacetime whereas a tilde
refers to the interal space.

For simplicity we take a maximal symmetric space as the internal space.
In short:
\begin{equation}
\Theta^{ab}=\kappa e^a\wedge e^b\quad {\rm with} \quad
\kappa={\rm constant}\,.    
\end{equation}
Hereafter, for concreteness, we make use of the hyperspheres as
maximal symmetric spaces, and set $\kappa=I/r^2$, where $r$ is the
radius of the sphere $S^N$.

The compactification on $S^N$ gives the four-dimensional action as
follows:
\begin{eqnarray}
I&=&K V_N \int d^4x\,
\hat{e}\left(\tilde{L}_m+m\tilde{L}_{m-1}\hat{R}\right.\nonumber \\
& &\left.+\frac{m(m-1)}{2}\tilde{L}_{m-2}
(\hat{R}_{\alpha\beta\gamma\delta}^2
-4\hat{R}_{\alpha\beta}^2+\hat{R}^2)\right)\,,
\label{14}
\end{eqnarray}
where $V_N$ is the volume of $S^N$.
After a simple combinatorial counting, we get
\begin{eqnarray}
\tilde{L}_m&=&\frac{N!}{(N-2m)!} (1/r^2)^m\,,\quad (N\ge 2m) \nonumber
\\
&=&0\,,\quad (N<2m)
\end{eqnarray}
in the case of $S^N$.
The first and second terms of the action (\ref{14}) correspond to the
cosmological term and the Einstein action respectively. We require here
$\tilde{L}_m=0$ and $\tilde{L}_{m-1}\ne 0$, which are fulfilled when
$D=4+N=2m+2$ or $2m+3$. We adopt here the case $D=4+N=2m+2$, because
this case can be regarded as a natural generalization from the Einstein
gravity in four dimensions, since a constant $K$ has dimension $({\rm
mass})^2$ and is the same as the inverse of the Newton constant in the
four-dimensional Einstein gravity.

We write the action with an appropriate normalization,
\begin{equation}
I=\int
d^{4+N}z\,e\left[\frac{1}{2N!m\kappa^2}L_m+L_{\rm matter}\right]\,,
\end{equation}
where $4+N=2m+2$, $L_{\rm matter}$, is the Lagrangian of
matter and $\kappa^2$ is a generalized Newton constant.
For convenience, we still use both $N$ and $m$
through this paper.

Next, we investigate cosmological solutions.
We assume the following metric in $4+N$ dimensions:
\begin{equation}
ds^2= - dt^2+R^2(t)(d\Omega_3)^2+ r^2(t)(d\Omega_N)^2\,,
\end{equation}
where $(d\Omega_{3(N)})^2$ is the line element of $3(N)$-dimensional
maximal symmetric space with a unit radius. As the previous case, We
take $S^N$ as extra spaces. Under these assumptions, the action is
written as
\begin{eqnarray}
I&\simeq&\int dt\,
(R^3r^N)\left[\frac{1}{2\kappa^2}\left\{\left(
\frac{1+\dot{r}^2}{r^2}\right)^{m-1}\left(-6\frac{\dot{R}^2}{R^2}+
\frac{k}{R^2}
\right)\right.\right.\nonumber \\
& &\qquad \qquad \qquad
-4(m-1)\frac{\dot{r}}{r}\left(\frac{\dot{R}}{R}\right)^3
\left(
\frac{1+\dot{r}^2}{r^2}\right)^{m-2}\nonumber \\
& &
-12(m-1)\left.\left.\frac{(rR)\dot{{}}}{R^3r^N}k\sum_{n=0}^{m-2}
\left(\begin{array}{c}m-2\\n\end{array}\right)\frac{1}{2n+1}
\dot{r}^{2n+1}\right\}-\rho(R, r) \right],
\end{eqnarray}
where $k=1, 0, -1$, corresponds to closed, flat and open
three-dimensional space, respectively. We have performed the partial
integration for the action to involve no second derivatives of scale
factors. Then this action leads to equations of motions with at most
second derivatives for $R$ or $r$. Taking variations with the metric,
we obtain the following equations of motions:
\begin{eqnarray}
& &3
\left(\frac{1+\dot{r}^2}{r^2}\right)^{m-1}
\left(\frac{\dot{R}}{R}\right)^2\nonumber \\
&+&6(m-1)
\left(\frac{1+\dot{r}^2}{r^2}\right)^{m-2}
\left\{
\left(\frac{\dot{r}}{r}\right)^2
\left(\frac{\dot{R}}{R}\right)^2
+
\frac{\dot{r}}{r}
\left(\frac{\dot{R}}{R}\right)^3
\right\}\nonumber\\
&+&4(m-1)(m-2)
\left(\frac{1+\dot{r}^2}{r^2}\right)^{m-3}
\left(\frac{\dot{r}}{r}\right)^3
\left(\frac{\dot{R}}{R}\right)^3
=\kappa^2\rho\,,
\label{19a}
\end{eqnarray}
\begin{eqnarray}
& &
\left(\frac{1+\dot{r}^2}{r^2}\right)^{m-1}
\left\{2\frac{\ddot{R}}{R}+\left(\frac{\dot{R}}{R}\right)^2\right\}
\nonumber
\\ &+&2(m-1)
\left(\frac{1+\dot{r}^2}{r^2}\right)^{m-2}
\left\{
\left(\frac{\dot{r}}{r}\right)^2
\left(\frac{\dot{R}}{R}\right)^2
+
\frac{\ddot{r}}{r}
\left(\frac{\dot{R}}{R}\right)^2
+
2\left(\frac{\ddot{r}}{r}+\frac{\ddot{R}}{R}\right)
\frac{\dot{r}}{r}\frac{\dot{R}}{R}
\right\}\nonumber\\
&+&4(m-1)(m-2)
\left(\frac{1+\dot{r}^2}{r^2}\right)^{m-3}
\frac{\ddot{r}}{r}
\left(\frac{\dot{r}}{r}\right)^2
\left(\frac{\dot{R}}{R}\right)^2
=-\kappa^2 p\,,
\label{19b}
\end{eqnarray}
\begin{eqnarray}
& &3
\left(\frac{1+\dot{r}^2}{r^2}\right)^{m-2}
\left\{\left(\frac{\ddot{R}}{R}+\frac{\ddot{r}}{r}
\right)\left(\frac{\dot{R}}{R}\right)^2+
\frac{\dot{r}}{r}\left(\frac{\dot{R}}{R}\right)^3
+2
\frac{\ddot{R}}{R}\frac{\dot{r}}{r}\frac{\dot{R}}{R}
\right\}\nonumber
\\ &+&2(m-2)
\left(\frac{1+\dot{r}^2}{r^2}\right)^{m-3}
\left\{
\left(\frac{\dot{r}}{r}\right)^3
\left(\frac{\dot{R}}{R}\right)^3
+3
\frac{\ddot{r}}{r}\frac{\dot{r}}{r}
\left(\frac{\dot{R}}{R}\right)^3
\right.\nonumber \\
& &\qquad\qquad\qquad\qquad\qquad\qquad+
\left.
3\left(\frac{\ddot{r}}{r}+\frac{\ddot{R}}{R}\right)
\left(\frac{\dot{r}}{r}\right)^2\left(\frac{\dot{R}}{R}\right)^2
\right\}\nonumber\\
&+&4(m-2)(m-3)
\left(\frac{1+\dot{r}^2}{r^2}\right)^{m-4}
\frac{\ddot{r}}{r}
\left(\frac{\dot{r}}{r}\right)^3
\left(\frac{\dot{R}}{R}\right)^3
=-\kappa^2 q\,,
\label{19c}
\end{eqnarray}
where $p=-(1/3R^2)(\partial/\partial R)(R^3\rho)$,
$q=-(1/Nr^{N-1})(\partial/\partial r)(r^N\rho)$.\cite{17} We wrote down
them only in the case $k=0$ for later use. We will examine here the
later stage of evolutions of scale factors. In such a case, we can
regard the matter as ``four-dimensional radiation''.\cite{6,18} In other
words, the equation of state is expressed as $\rho=3p$ ($q=0$). This
ansatz should be valid in the region $1/R<T<1/r$, where $T$ is the
temperature.\cite{18}

There is an approximate Friedmann (Tolemann) solution with the
relatively slow time variation of $r$, we find that in the case $k=0$,
\begin{eqnarray}
R&\sim&R_0 t^{1/2}(1+ O((m-1) r_0^2 t^{2(\beta-1)}))\,,
\label{20a}\\
r&\sim&r_0 t^{\beta}(1+ O((m-1) r_0^2 t^{2(\beta-1)}))\,,
\label{20b}
\end{eqnarray}
where $\beta=(3-\sqrt{13})/4\sim -0.151$. These results
may be compared with the solution in the usual
five-dimensional Kaluza-Klein cosmology with
``four-dimensional radiation''.\cite{19} Equation (\ref{20b})
shows that the scale of the compact space shrinks
asymptotically more slowly in our model.

However, our model has a distinct property. The third equation of motion
(\ref{19c}) shows that $\dot{r}=0$ is forbidden even if we intended to
set it by hand in the case $q=0$, it means that the particle creation by
quantum effects\cite{20} plays an important role because the leading
term on the left-hand side of the equation involves the fourth-order of
time denvatives. This problem will be examined in the near future.

If we want to investigate whether the Kaluza-Klein inflation takes
place or not, we should take into consideration highly
nonlinear effects of the differential equations and need to perform
numerical  calculations  carefully. The  Regge calculus\cite{21} may be
suitable for the calculation.

Next, we should remark on the static solutions including dimensional
reduction.\cite{22} In our model, the metric of
(Schwarzschild solution)$\times$(constant
$S^N$) is not a solution. The study of properties of
the static solutions in our theory is currently in progress.
 
Finally, we will comment on quantum nature of our theory. In flat
space-time, it is known that the extended Gauss-Bonnet terms contain
only the graviton interaction vertex.\cite{14} However, contents of
these terms is still unknown in curved space-time. While we need the
principle that forbids the existence of the cosmological term,
the Einstein-Hilbert term and all others unrequired, we should consider
the supersymmetric extension of our model. Therefore, it is important
to investigate quantum nature of the theory, as well as contributions
of other matters. Particularly, to supersymmetrize the extended
Gauss-Bonnet term, we need to introduce antisymetric
tensors,\cite{23} which also affects cosmological solutions.

We would like to thank H. Sugawara, M. Yoshimura and M. Kobayashi for
their warm hospitality at KEK, where most part of this work was done. We
wish to thank M. Hosoda for critical reading of the manuscript.

%%%%%%%%%%%%%%%%%%%%%%%%%%%%%%%%%%%%%%%%%%%%%% 

%%%%%%%%%%%%%%%%%%%%%%%%%%%%%%%%%%%%%%%%%%%%%
\end{document}